\documentclass[aps,prc,twocolumn,superscriptaddress,floatfix,showpacs]{revtex4}

\usepackage{graphicx}

  \newcommand*{\Journal}[4]{#1 \textbf{#2}, #3 (#4)}
  \newcommand*{\EPJC}{Eur. Phys. J. C}
  \newcommand*{\PL}{Phys. Lett.}
  \newcommand*{\ZPC}{Z. Phys. C}
  \newcommand*{\JPG}{J. Phys. G}
  \newcommand*{\APPB}{Acta Phys. Pol. B}
  \newcommand*{\NIMA}{Nucl. Instrum. Methods A}
  \newcommand*{\PRL}{Phys. Rev. Lett.}
  \newcommand*{\CPC}{Comput. Phys. Commun.}
  \newcommand*{\PRD}{Phys. Rev. D}

  \newcommand*{\agev}{$A$~GeV}
  \newcommand*{\dzero}{\ensuremath{\text{D}^0}}
  \newcommand*{\dbzero}{\ensuremath{\bar{\text{D}}^{0}}}
  \newcommand*{\kmin}{\ensuremath{\text{K}^-}}
  \newcommand*{\kplus}{\ensuremath{\text{K}^+}}
  \newcommand*{\pimin}{\ensuremath{\pi^-}}
  \newcommand*{\piplus}{\ensuremath{\pi^+}}
  \newcommand*{\rarrow}{\ensuremath{\rightarrow}}
  \newcommand*{\tm}{\ensuremath{\times}}
  \newcommand*{\dedx}{\ensuremath{\text{d}E/\text{d}x}}
  \newcommand*{\De}{\Delta}
  \newcommand*{\ga}{\gamma}
  \newcommand*{\Ga}{\Gamma}
  \newcommand*{\si}{\sigma}
  \newcommand*{\peetee}{\ensuremath{p_{\text{T}}}}
  \newcommand*{\ndee}{\ensuremath{N}}
  \newcommand*{\npart}{\ensuremath{N_{\text{part}}}}
  \newcommand*{\der}{\ensuremath{\text{d}}}
  \newcommand*{\beq}[1]{\begin{equation}\label{#1}}
  \newcommand*{\eeq}{\end{equation}}
  \newcommand*{\bea}[1]{\begin{eqnarray}\label{#1}}
  \newcommand*{\eea}{\end{eqnarray}}
  \newcommand*{\Fi}[1]{Fig.~\ref{#1}}
  \newcommand*{\Eq}[1]{Eq.~(\ref{#1})}

\begin{document}


\title{ 
  Upper Limit of D$^0$ Production in Central Pb-Pb Collisions\\
  at 158\agev
  }


\affiliation{NIKHEF, Amsterdam, Netherlands.}  
\affiliation{Department of Physics, University of Athens, Athens, Greece.}
\affiliation{Comenius University, Bratislava, Slovakia.}
\affiliation{KFKI Research Institute for Particle and Nuclear Physics,
             Budapest, Hungary.}
\affiliation{MIT, Cambridge, USA.}
\affiliation{Institute of Nuclear Physics, Cracow, Poland.}
\affiliation{Gesellschaft f\"{u}r Schwerionenforschung (GSI),
             Darmstadt, Germany.} 
\affiliation{Joint Institute for Nuclear Research, Dubna, Russia.}
\affiliation{Fachbereich Physik der Universit\"{a}t, Frankfurt,
             Germany.}
\affiliation{CERN, Geneva, Switzerland.}
\affiliation{Gesellschaft f\"{u}r Schwerionenforschung (GSI),
             Darmstadt, Germany.}
\affiliation{Institute of Physics \'Swi{\,e}tokrzyska Academy, Kielce,
             Poland.} 
\affiliation{Fachbereich Physik der Universit\"{a}t, Marburg, Germany.}
\affiliation{Max-Planck-Institut f\"{u}r Physik, Munich, Germany.}
\affiliation{Institute of Particle and Nuclear Physics, Charles
             University, Prague, Czech Republic.}
\affiliation{Department of Physics, Pusan National University, Pusan,
             Republic of Korea.} 
\affiliation{Nuclear Physics Laboratory, University of Washington,
             Seattle, WA, USA.} 
\affiliation{Atomic Physics Department, Sofia University St.~Kliment
             Ohridski, Sofia, Bulgaria.}
\affiliation{Institute for Nuclear Research and Nuclear Energy, Sofia,
             Bulgaria.}  
\affiliation{Institute for Nuclear Studies, Warsaw, Poland.}
\affiliation{Institute for Experimental Physics, University of Warsaw,
             Warsaw, Poland.} 
\affiliation{Rudjer Boskovic Institute, Zagreb, Croatia.}


\author{C.~Alt}
\affiliation{Fachbereich Physik der Universit\"{a}t, Frankfurt,
             Germany.}
\author{T.~Anticic} 
\affiliation{Rudjer Boskovic Institute, Zagreb, Croatia.}
\author{B.~Baatar}
\affiliation{Joint Institute for Nuclear Research, Dubna, Russia.}
\author{D.~Barna}
\affiliation{KFKI Research Institute for Particle and Nuclear Physics,
             Budapest, Hungary.} 
\author{J.~Bartke}
\affiliation{Institute of Nuclear Physics, Cracow, Poland.}
\author{L.~Betev}
\affiliation{CERN, Geneva, Switzerland.}
\author{H.~Bia{\l}\-kowska} 
\affiliation{Institute for Nuclear Studies, Warsaw, Poland.}
\author{C.~Blume}
\affiliation{Fachbereich Physik der Universit\"{a}t, Frankfurt,
             Germany.}
\author{B.~Boimska}
\affiliation{Institute for Nuclear Studies, Warsaw, Poland.}
\author{M.~Botje}
\affiliation{NIKHEF, Amsterdam, Netherlands.}
\author{J.~Bracinik}
\affiliation{Comenius University, Bratislava, Slovakia.}
\author{R.~Bramm}
\affiliation{Fachbereich Physik der Universit\"{a}t, Frankfurt,
             Germany.}
\author{P.~Bun\v{c}i\'{c}}
\affiliation{CERN, Geneva, Switzerland.}
\author{V.~Cerny}
\affiliation{Comenius University, Bratislava, Slovakia.}
\author{P.~Christakoglou}
\affiliation{Department of Physics, University of Athens, Athens, Greece.}
\author{O.~Chvala}
\affiliation{Institute of Particle and Nuclear Physics, Charles
             University, Prague, Czech Republic.} 
\author{J.G.~Cramer}
\affiliation{Nuclear Physics Laboratory, University of Washington,
             Seattle, WA, USA.} 
\author{P.~Csat\'{o}} 
\affiliation{KFKI Research Institute for Particle and Nuclear Physics,
             Budapest, Hungary.}
\author{P.~Dinkelaker}
\affiliation{Fachbereich Physik der Universit\"{a}t, Frankfurt,
             Germany.}
\author{V.~Eckardt}
\affiliation{Max-Planck-Institut f\"{u}r Physik, Munich, Germany.}
\author{D.~Flierl}
\affiliation{Fachbereich Physik der Universit\"{a}t, Frankfurt,
             Germany.} 
\author{Z.~Fodor}
\affiliation{KFKI Research Institute for Particle and Nuclear Physics,
             Budapest, Hungary.} 
\author{P.~Foka}
\affiliation{Gesellschaft f\"{u}r Schwerionenforschung (GSI),
             Darmstadt, Germany.} 
\author{V.~Friese}
\affiliation{Gesellschaft f\"{u}r Schwerionenforschung (GSI),
             Darmstadt, Germany.} 
\author{J.~G\'{a}l}
\affiliation{KFKI Research Institute for Particle and Nuclear Physics,
             Budapest, Hungary.} 
\author{M.~Ga\'zdzicki}
\affiliation{Fachbereich Physik der Universit\"{a}t, Frankfurt,
             Germany.}
\affiliation{Institute of Physics \'Swi{\,e}tokrzyska Academy, Kielce,
             Poland.}
\author{V.~Genchev}
\affiliation{Institute for Nuclear Research and Nuclear Energy, Sofia,
             Bulgaria.} 
\author{G.~Georgopoulos}
\affiliation{Department of Physics, University of Athens, Athens,
             Greece.}
\author{E.~G{\l}adysz}
\affiliation{Institute of Nuclear Physics, Cracow, Poland.}
\author{K.~Grebieszkow}
\affiliation{Institute for Experimental Physics, University of Warsaw,
             Warsaw, Poland.} 
\author{S.~Hegyi}
\affiliation{KFKI Research Institute for Particle and Nuclear Physics,
             Budapest, Hungary.} 
\author{C.~H\"{o}hne}
\affiliation{Gesellschaft f\"{u}r Schwerionenforschung (GSI),
             Darmstadt, Germany.}
\author{K.~Kadija}
\affiliation{Rudjer Boskovic Institute, Zagreb, Croatia.}
\author{A.~Karev}
\affiliation{Max-Planck-Institut f\"{u}r Physik, Munich, Germany.}
\author{M.~Kliemant}
\affiliation{Fachbereich Physik der Universit\"{a}t, Frankfurt,
             Germany.}
\author{S.~Kniege}
\affiliation{Fachbereich Physik der Universit\"{a}t, Frankfurt,
             Germany.}
\author{V.I.~Kolesnikov}
\affiliation{Joint Institute for Nuclear Research, Dubna, Russia.}
\author{E.~Kornas}
\affiliation{Institute of Nuclear Physics, Cracow, Poland.}
\author{R.~Korus}
\affiliation{Institute of Physics \'Swi{\,e}tokrzyska Academy, Kielce,
             Poland.} 
\author{M.~Kowalski}
\affiliation{Institute of Nuclear Physics, Cracow, Poland.}
\author{I.~Kraus}
\affiliation{Gesellschaft f\"{u}r Schwerionenforschung (GSI),
             Darmstadt, Germany.} 
\author{M.~Kreps}
\affiliation{Comenius University, Bratislava, Slovakia.}
\author{M.~van~Leeuwen}
\affiliation{NIKHEF, Amsterdam, Netherlands.}
\author{P.~L\'{e}vai}
\affiliation{KFKI Research Institute for Particle and Nuclear Physics,
             Budapest, Hungary.} 
\author{L.~Litov}
\affiliation{Atomic Physics Department, Sofia University St.~Kliment
             Ohridski, Sofia, Bulgaria.}
\author{B.~Lungwitz}
\affiliation{Fachbereich Physik der Universit\"{a}t, Frankfurt,
             Germany.}  
\author{M.~Makariev}
\affiliation{Atomic Physics Department, Sofia University St.~Kliment
             Ohridski, Sofia, Bulgaria.} 
\author{A.I.~Malakhov}
\affiliation{Joint Institute for Nuclear Research, Dubna, Russia.}
\author{M.~Mateev}
\affiliation{Atomic Physics Department, Sofia University St.~Kliment
             Ohridski, Sofia, Bulgaria.} 
\author{G.L.~Melkumov}
\affiliation{Joint Institute for Nuclear Research, Dubna, Russia.}
\author{A.~Mischke}
\affiliation{NIKHEF, Amsterdam, Netherlands.}
\author{M.~Mitrovski}
\affiliation{Fachbereich Physik der Universit\"{a}t, Frankfurt,
             Germany.} 
\author{J.~Moln\'{a}r}
\affiliation{KFKI Research Institute for Particle and Nuclear Physics,
             Budapest, Hungary.} 
\author{St.~Mr\'owczy\'nski}
\affiliation{Institute of Physics \'Swi{\,e}tokrzyska Academy, Kielce,
             Poland.}
\author{V.~Nicolic}
\affiliation{Rudjer Boskovic Institute, Zagreb, Croatia.}
\author{G.~P\'{a}lla}
\affiliation{KFKI Research Institute for Particle and Nuclear Physics,
             Budapest, Hungary.} 
\author{A.D.~Panagiotou}
\affiliation{Department of Physics, University of Athens, Athens,
             Greece.} 
\author{D.~Panayotov}
\affiliation{Atomic Physics Department, Sofia University St.~Kliment
             Ohridski, Sofia, Bulgaria.} 
\author{A.~Petridis}
\affiliation{Department of Physics, University of Athens, Athens,
             Greece.} 
\author{M.~Pikna}
\affiliation{Comenius University, Bratislava, Slovakia.}
\author{D.~Prindle}
\affiliation{Nuclear Physics Laboratory, University of Washington,
             Seattle, WA, USA.}
\author{F.~P\"{u}hlhofer}
\affiliation{Fachbereich Physik der Universit\"{a}t, Marburg, Germany.}
\author{R.~Renfordt}
\affiliation{Fachbereich Physik der Universit\"{a}t, Frankfurt,
             Germany.} 
\author{C.~Roland}
\affiliation{MIT, Cambridge, USA.}
\author{G.~Roland}
\affiliation{MIT, Cambridge, USA.}
\author{M.~Rybczy\'nski}
\affiliation{Institute of Physics \'Swi{\,e}tokrzyska Academy, Kielce,
             Poland.}
\author{A.~Rybicki}
\affiliation{Institute of Nuclear Physics, Cracow, Poland.}
\affiliation{CERN, Geneva, Switzerland.}
\author{A.~Sandoval}
\affiliation{Gesellschaft f\"{u}r Schwerionenforschung (GSI),
             Darmstadt, Germany.} 
\author{N.~Schmitz}
\affiliation{Max-Planck-Institut f\"{u}r Physik, Munich, Germany.}
\author{T.~Schuster}
\affiliation{Fachbereich Physik der Universit\"{a}t, Frankfurt,
             Germany.}
\author{P.~Seyboth}
\affiliation{Max-Planck-Institut f\"{u}r Physik, Munich, Germany.}
\author{F.~Sikl\'{e}r}
\affiliation{KFKI Research Institute for Particle and Nuclear Physics,
             Budapest, Hungary.} 
\author{B.~Sitar}
\affiliation{Comenius University, Bratislava, Slovakia.}
\author{E.~Skrzypczak}
\affiliation{Institute for Experimental Physics, University of Warsaw,
             Warsaw, Poland.} 
\author{G.~Stefanek}
\affiliation{Institute of Physics \'Swi{\,e}tokrzyska Academy, Kielce,
             Poland.}
\author{R.~Stock}
\affiliation{Fachbereich Physik der Universit\"{a}t, Frankfurt,
             Germany.}
\author{H.~Str\"{o}bele}
\affiliation{Fachbereich Physik der Universit\"{a}t, Frankfurt,
             Germany.}
\author{T.~Susa}
\affiliation{Rudjer Boskovic Institute, Zagreb, Croatia.}
\author{I.~Szentp\'{e}tery}
\affiliation{KFKI Research Institute for Particle and Nuclear Physics,
             Budapest, Hungary.} 
\author{J.~Sziklai}
\affiliation{KFKI Research Institute for Particle and Nuclear Physics,
             Budapest, Hungary.}
\author{P.~Szymanski}
\affiliation{CERN, Geneva, Switzerland.}
\affiliation{Institute for Nuclear Studies, Warsaw, Poland.}
\author{V.~Trubnikov}
\affiliation{Institute for Nuclear Studies, Warsaw, Poland.}
\author{D.~Varga}
\affiliation{KFKI Research Institute for Particle and Nuclear Physics,
             Budapest, Hungary.}
\affiliation{CERN, Geneva, Switzerland.} 
\author{M.~Vassiliou}
\affiliation{Department of Physics, University of Athens, Athens,
             Greece.}
\author{G.I.~Veres}
\affiliation{KFKI Research Institute for Particle and Nuclear Physics,
             Budapest, Hungary.} 
\affiliation{MIT, Cambridge, USA.}
\author{G.~Vesztergombi}
\affiliation{KFKI Research Institute for Particle and Nuclear Physics,
             Budapest, Hungary.}
\author{D.~Vrani\'{c}}
\affiliation{Gesellschaft f\"{u}r Schwerionenforschung (GSI),
             Darmstadt, Germany.} 
\author{A.~Wetzler}
\affiliation{Fachbereich Physik der Universit\"{a}t, Frankfurt,
             Germany.}
\author{Z.~W{\l}odarczyk}
\affiliation{Institute of Physics \'Swi{\,e}tokrzyska Academy, Kielce,
             Poland.}
\author{I.K.~Yoo}
\affiliation{Department of Physics, Pusan National University, Pusan,
             Republic of Korea.} 
\author{J.~Zim\'{a}nyi}
\affiliation{KFKI Research Institute for Particle and Nuclear Physics,
             Budapest, Hungary.} 

\collaboration{The NA49 collaboration}
\noaffiliation

\date{\today}


\begin{abstract}
Results are presented from a search for the decays
$\dzero \rarrow \kmin \piplus$ and $\dbzero \rarrow \kplus \pimin$ in
a sample of $3.8 \tm 10^6$ central Pb-Pb events collected with a
beam energy of 158\agev\ by NA49 at the CERN SPS. No signal is
observed. An upper limit on \dzero\ production is derived and compared
to predictions from several models.
\end{abstract}

\pacs{25.75.Dw}

\maketitle


The measurement of open charm production in heavy-ion interactions is
of considerable interest because charm, due to its large mass, is
predominantly created at the early stage of the collision when the
energy density is large.  Because of the hard scale involved,
perturbative QCD (pQCD) calculations can serve as a baseline for the
study of the production mechanisms and the dynamical evolution of
charm in these collisions.

At present, no direct measurement exists of open charm production in
heavy-ion interactions at SPS energies.  The NA38/50 experiment has,
however, observed a significant enhancement of di-muon production in
the intermediate mass range of 1.5--2.5~GeV, compared to di-muon
yields expected from the Drell-Yan continuum and semi-leptonic charm
decays~\cite{ref:na3850}.  The origin of this enhancement is presently
not clear but can be explained by assuming that open charm production
in central Pb-Pb collisions is about a factor of 3.5 times larger than
predicted by pQCD. This enhancement is currently under investigation
by the NA60 experiment~\cite{ref:na60}.

A variety of models give very different estimates for the open charm
yields at the SPS. For instance, for central Pb-Pb interactions at
158\agev\ beam energy, a pQCD calculation based on Pythia predicts a
yield per event of $N(\dzero+\dbzero) = 0.21$ (the centrality is here
characterized by the number of participant nucleons $\npart =
400$)~\cite{ref:pbmcharm}.  In~\cite{ref:gorenstein}, a yield of
0.5--0.6 $c\bar{c}$ quark pairs is calculated, based on the J/$\Psi$
yield measured by NA50 and the statistical coalescence model ($\npart
= 360$). This translates into $N(\dzero+\dbzero) \approx 0.4$ if one
assumes that about one third of the $c\bar{c}$ hadronize into \dzero\
and \dbzero, like in p-p interactions~\cite{ref:lebc}. The ALCOR
hadronization model~\cite{ref:alcor} gives a much larger estimate of
$N(\dzero+\dbzero) = 2.4$ ($\npart \approx 350$).  An even larger
yield of $N(\dzero+\dbzero) \approx 6$ is predicted by the statistical
model of the early stage (SMES) which assumes charm equilibration in a
deconfined quark-gluon plasma (QGP) at the early stage of the Pb-Pb
interaction ($\npart = 360$)~\cite{ref:smes}.

To discriminate between the different model predictions and to
possibly identify the origin of the di-muon enhancement seen by
NA38/50, we have performed a search for open charm, using invariant
mass reconstruction, in a large data sample of about four million
central Pb-Pb events collected at 158\agev\ beam energy.

The NA49 detector~\cite{ref:na49nimpaper} is a large acceptance
fixed-target hadron spectrometer at the CERN SPS. Tracking is
performed by four large-volume TPCs. Two of these are placed one
behind the other inside two super-conducting dipole magnets (vertex
TPCs). The two other (main) TPCs are placed downstream of the
magnets left and right of the beam line. These main TPCs increase the
lever arm of the track reconstruction and are optimized for particle
identification through a measurement of the specific energy loss
(\dedx) with a relative resolution of about 4\%. The combined TPCs
provide an accurate measurement of the particle momenta with a
resolution of $\De p/p^2 \approx 3 \tm 10^{-5}$~(GeV/$c$)$^{-1}$.
Centrality selection is based on a measurement of the energy deposited
by the projectile spectator nucleons in a forward calorimeter.

To measure rare particles like the $\Omega$~\cite{ref:omegapap} and to
search for open charm a large sample of central Pb-Pb events was taken
in the year 2000 with a beam energy of 158\agev. In this run $3 \times
10^6$ events were collected with a centrality selection of 23.5\% of
the inelastic cross-section ($\npart = 262$).  Also included in the
present analysis is a 1996 data set of $8 \times 10^5$ Pb-Pb events,
taken at the same beam energy but with a 10\% centrality selection
($\npart = 335$). To increase the data acquisition speed and decrease
the data volume only every second time-sample of the TPCs was read out
during the 2000 run (256 instead of 512 time-samples). The reduced
sampling did not significantly affect the track reconstruction nor the
\dedx\ measurement.

The \dzero\ were reconstructed via their charged particle decays
$\dzero \rarrow \kmin \piplus$ and $\dbzero \rarrow \kplus \pimin$
(4\% branching ratio). Because the secondary vertex resolution of
about 1~cm is not sufficient to detect the decay vertex ($\ga c \tau
\approx 1$~mm), the \dzero\ candidates were identified by selecting a
window around the \dzero\ mass in the invariant mass spectrum of the
daughter particles. With a multiplicity of approximately 1400
reconstructed charged tracks, about $5 \tm 10^5$ entries for each
event were made in each of the \dzero\ and \dbzero\ invariant mass
spectra leading to a large combinatorial background.  Because of the
large multiplicities it was not possible to measure the 3-particle
decay $\text{D}^* \rarrow \text{K}\pi\pi$ even though the background in
this channel is suppressed by kinematic constraints.

Events for which the primary vertex could not be determined were
discarded from the analysis. Several track quality
cuts~\cite{ref:marcothesis} were applied to remove non-vertex or badly
reconstructed tracks.  The remaining sample was sub-divided into two
classes.  The tracks in the first class have sufficient track length
in the main TPCs and low enough momentum so that a significant
enrichment of the kaon content could be achieved by suitable cuts on
\dedx. For tracks in the second class this kaon identification was not
possible. Loose cuts on \dedx\ ($2\si$ around the Bethe-Bloch curve)
were applied on the tracks in the first sample to minimize the loss of
kaons (and \dzero).  The identified kaon tracks were then combined
with all oppositely charged tracks and the invariant mass of the pair
was calculated assuming that the associated track was a pion. In the
second class (without \dedx) the invariant mass of the \dzero\
(\dbzero) candidate was calculated for all pairs of oppositely charged
tracks assuming that the positive track was a pion (kaon) and the
negative track a kaon (pion). The invariant mass distributions
obtained from the \dzero\ and \dbzero\ samples with and without kaon
identification (corrected for acceptance and efficiency, see below)
are shown by the open histograms in \Fi{fig:d0fig1}.
%
\begin{figure}
\includegraphics[width=1.0\linewidth]{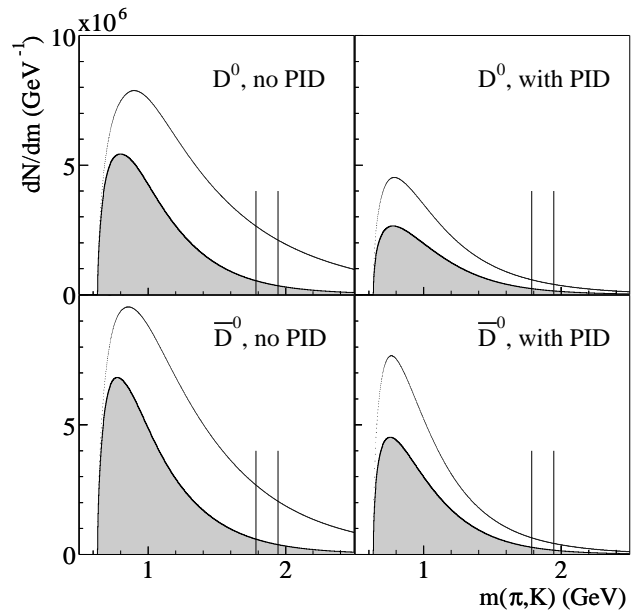}
\caption{\label{fig:d0fig1} 
  Invariant mass distributions of \dzero\ (top) and \dbzero\
  candidates (bottom) from track samples without (left) and with
  (right) kaon identification. The open (shaded) histograms are before
  (after) applying the decay angle cuts described in the text. The
  vertical lines indicate the \dzero\ mass window used in the
  analysis.  The distributions are corrected for acceptance and
  branching ratio.  }
\end{figure}

To further reduce the combinatorial background, decay angle cuts were
applied as follows. For each \dzero\ candidate the polar angle
$\theta$ and the azimuthal angle $\phi$ of the kaon track were
calculated in the rest-frame of the \dzero. Here $\theta$ is the angle
between the beam direction and the kaon flight direction and $\phi$
the angle between the kaon and the flight direction of the \dzero\ in
the plane perpendicular to the beam.  In the left-hand side plot of
\Fi{fig:d0fig2} is shown the distribution of decay angles from
simulated \dzero\ decays (see below).  The distribution from real
events (almost entirely background) is shown in the right-hand side
plot. It is clear from this figure that the signal distribution is
approximately flat while the background distribution peaks at large
values of $|\cos \theta|$.
%
\begin{figure}
\includegraphics[width=1.0\linewidth]{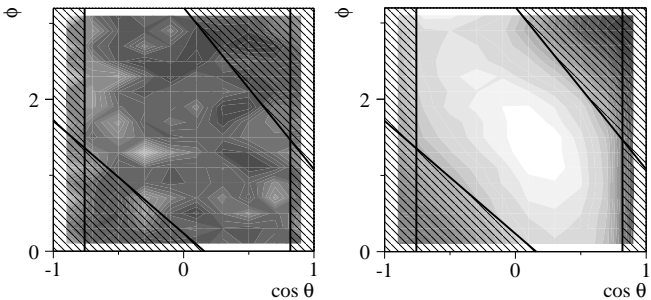}
\caption{\label{fig:d0fig2} 
  The reconstructed decay angle distributions (see text) of simulated
  \dzero\ decays (left) and of \dzero\ candidates from real events
  (right). In both plots the \dzero\ candidates have a reconstructed
  mass within 50~MeV of the nominal \dzero\ mass and a reconstructed
  \peetee\ of 800--1200~MeV. A darker shade corresponds to more
  entries in the plots. The full lines indicate the cuts used in 
  the analysis to discard the hatched regions.
  }
\end{figure}
%
Cuts, like those shown in the figure, were optimized to maximize the
significance ($= \mbox{signal}/\sqrt{\mbox{background}}$) of the
measurement.  Because the decay angle distribution depends on the
transverse momentum (\peetee) of the \dzero\ and is different for the
samples with and without particle identification, separate cuts were
determined, for each of the two samples, in five \peetee\ bins of
400~MeV width. The decay angle cuts reduced the background by a factor
of about 3 (10) in the sample with (without) particle
identification. (The signal is reduced by 30--40\%.)  Cuts on other
kinematic variables like rapidity ($y$) and \peetee\ were investigated
but were found to be ineffective in the separation of signal and
background~\cite{ref:marcothesis}. The invariant mass distributions
after the decay angle cuts and corrected for acceptance and efficiency
(see below) are shown by the shaded histograms in \Fi{fig:d0fig1}.

To determine acceptance, efficiency and mass resolution a Monte Carlo
sample of \dzero\ and \dbzero\ mesons was generated with a Gaussian
distribution in $y$ ($\si_y = 0.6$) and an exponential distribution in
transverse mass (300~MeV inverse slope parameter).  The \dzero\
(\dbzero) and their decay particles were transported through the NA49
detector geometry using GEANT 3.21~\cite{ref:geant}, followed by a
detailed simulation of the TPC response using dedicated NA49 software.
The simulated raw data were added to real events and subjected to the
same reconstruction procedure as the experimental data. The acceptance
was calculated in bins of $y$ and \peetee\ as the fraction of \dzero\
(\dbzero) which are geometrically accepted, survive the reconstruction
procedure and pass the analysis cuts. 
The experimental acceptance covers the range
$\peetee > 0$ and $-1 \alt y \alt 1.6$ and is found to be, on average,
8.4 (12.0)\% for the sample with (without) kaon identification. It was
verified that the amount of accepted particles varied by only 10--20\%
if reasonable alternatives\ (e.g.\ from Pythia~\cite{ref:pythia}) were
chosen for the kinematic distribution of the \dzero. The invariant
mass distributions shown in \Fi{fig:d0fig1} are divided by the
acceptance and by the branching ratio for $\text{D} \rarrow \pi
\text{K}$ decay.

The simulated data served to determine the shape of the invariant
mass distribution of reconstructed \dzero\ as shown in \Fi{fig:d0fig3}.
%
\begin{figure}
\includegraphics[width=1.0\linewidth]{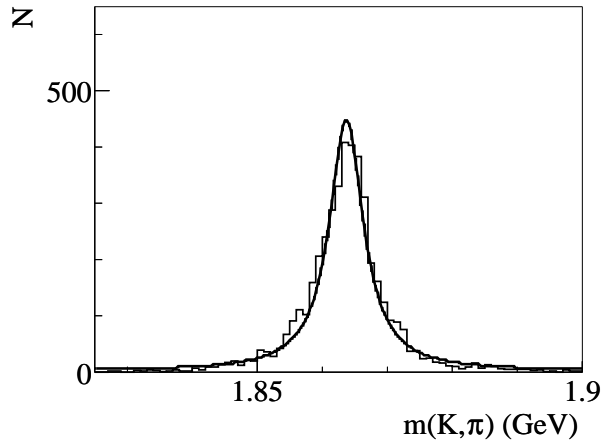}
\caption{\label{fig:d0fig3}
  Invariant mass distribution of simulated \dzero\ decays embedded in
  real data events. The full curve shows the Cauchy parameterization
  described in the text.
  }
\end{figure}
%
The shape can be well described by a Cauchy distribution (curve in
\Fi{fig:d0fig3})
\beq{eq:signalshape}
  \frac{\der n}{\der m} = 
  \frac{\ndee}{2\pi}\ \frac{\Ga}{(m-m_0)^2 + (\Ga/2)^2},
\eeq
where $\ndee$ is the total \dzero\ yield per event, $m_0$ the \dzero\
mass and $\Ga$ the width of the distribution. This width is almost
entirely determined by the detector resolution and is found to be $\Ga
= 6.2$~MeV with the mass of the \dzero\ set to $m_0 =
1864.5$~MeV~\cite{ref:pdg}.

The invariant mass spectra shown in \Fi{fig:d0fig1} were fitted (by
$\chi^2$ minimization) in a region of $\pm 90$~MeV around the nominal
\dzero\ mass to the sum of a signal distribution, \Eq{eq:signalshape},
and a fourth order polynomial describing the background.  The position
$m_0$ and width $\Ga$ of the signal distribution were kept fixed to
the values given above while the normalization $\ndee$ was left a free
parameter in the fit.  This fit results in yields (per event) of
$N(\dzero) = -0.41 \pm 0.51$ and $N(\dbzero) = 0.05 \pm 0.54$, where
the errors are statistical only. In \Fi{fig:d0fig4} is shown the
invariant mass distribution of the $\dzero + \dbzero$ candidates after
background subtraction. Clearly no signal is observed. The fit gave
for the total yield a value of $N(\dzero+\dbzero) = -0.36 \pm 0.74$
per event (full line in \Fi{fig:d0fig4}).

An upper limit for the number of \dzero\ per event is estimated in a
Bayesian approach~\cite{ref:sivia}. Here the likelihood $P({\rm
data}|\ndee)$ (i.e.\ the conditional probability density distribution
of the data, given $\ndee$ \dzero\ per event) is parameterized as a
Gaussian
\beq{eq:likelihood} 
  P(\text{data}|\ndee) = \frac{1}{\si \sqrt{2 \pi}} \exp \left[ -
  \frac{(\ndee - \mu)^2}{2 \si^2}\right] \equiv g(\ndee;\mu,\si)
\eeq
with mean $\mu = -0.36$ and width $\si = 0.74$ as obtained from the
$\chi^2$ fit. Using Bayes' theorem the posterior distribution
$P(\ndee|{\rm data})$ is calculated by multiplying the likelihood with
an assumed prior probability distribution of $\ndee$ which is taken
here to be zero for $\ndee < 0$ and uniform for $\ndee \geq 0$.
%
\begin{figure}
\includegraphics[width=1.0\linewidth]{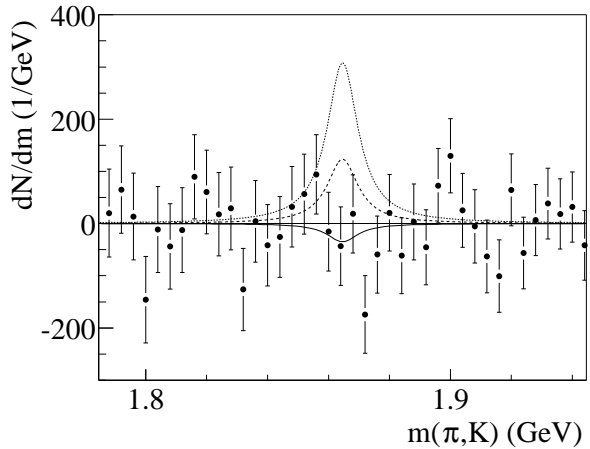}
\caption{\label{fig:d0fig4} 
  Invariant mass distribution of the $\dzero + \dbzero$ candidates
  after background subtraction. The errors are statistical only. The
  full curve shows the fit to the signal distribution described in the
  text, the dashed curve the expectation from ALCOR~\cite{ref:alcor}
  and the dotted curve that from SMES~\cite{ref:smes}.  }
\end{figure}
%
This prior distribution forces $\ndee$ to be positive, as it should
be.  Integration of the posterior distribution gives for the
confidence level
\beq{eq:conflevel}
  \text{CL} \equiv \int_0^{M} P(\ndee|\text{data})\;\der \ndee =
           \frac{\int_0^{M} g(\ndee;\mu,\si)\;\der \ndee}
           {\int_0^{\infty}g(\ndee;\mu,\si)\;\der \ndee},
\eeq
where $M$ is the upper limit of $\ndee$ corresponding to the
confidence level CL.  The denominator on the right-hand side of
\Eq{eq:conflevel} accounts for the proper normalization of
$P(\ndee|\text{data})$.  Using the fitted values of $\mu$ and $\si$,
the upper limit for the total yield is found to be $M(\dzero+\dbzero)
= 1.5$ per event at 98\% CL.

Because no \dzero\ signal has been observed it is not possible to
directly verify the Monte Carlo prediction of the signal shape. To
investigate the sensitivity of the upper limit to the width of the
mass peak the fits were repeated with $\Ga = 12.4$~MeV.  This resulted
in $N(\dzero) = -0.46 \pm 0.85$, $N(\dbzero) = -0.22 \pm 0.90$,
$N(\dzero+\dbzero) = -0.7 \pm 1.2$ and an upper limit of
$M(\dzero+\dbzero) = 2.4$ per event at 98\% CL. We remark that
increasing the Monte Carlo estimate of the width by a factor of two
should be considered a very generous error on $\Ga$.

Taking as a standard for comparison the pQCD estimate of
$N(\dzero+\dbzero) = 0.21$ mentioned in the introduction, we conclude
that an enhancement of charm production by more than a factor of 5--10
at the SPS is very unlikely.  Due to the large combinatorial
background it is not possible to confirm, nor exclude, a charm
enhancement by a factor of three allowed by the NA38/50 measurement.
However, the \dzero\ upper limit from this analysis is only marginally
compatible with the yield estimated by the ALCOR model (dashed curve
in \Fi{fig:d0fig4}) and clearly incompatible with the equilibrium
yield of charm in a QGP as predicted by the SMES (dotted curve in
\Fi{fig:d0fig4}).  The latter observation does not necessarily
exclude QGP formation at SPS energies provided that the QGP life time
is shorter than the equilibration time of charm.

\begin{acknowledgments}
This work was supported by the US Department of Energy
Grant DE-FG03-97ER41020/A000,
the Bundesministerium fur Bildung und Forschung, Germany, 
the Virtual Institute VI-146 of Helmholtz Gemeinschaft, Germany,
the Polish State Committee for Scientific Research (2 P03B 130 23,
SPB/CERN/P-03/Dz 446/2002-2004, 2 P03B 04123),  
the Hungarian Scientific Research Foundation (T032648, T032293, T043514),
the Hungarian National Science Foundation, OTKA, (F034707),
the Polish-German Foundation, and the Korea Research Foundation Grant
(KRF-2003-070-C00015). 
\end{acknowledgments}



\end{document}